\documentclass{article}

\usepackage{PRIMEarxiv}

\usepackage[utf8]{inputenc} 
\usepackage[T1]{fontenc}    
\usepackage{hyperref}       
\usepackage{url}            
\usepackage{booktabs}       
\usepackage{amsfonts}       
\usepackage{nicefrac}       
\usepackage{microtype}      
\usepackage{lipsum}
\usepackage{fancyhdr}       
\usepackage{graphicx}       
\graphicspath{{media/}}     

\title{From RISC-V Cores to Neuromorphic Arrays: A Tutorial on Building Scalable Digital Neuromorphic Processors
}

\author{
  Amirreza Yousefzadeh\\
  University of Twente, The Netherlands \\
  \texttt{a.yousefzadeh@utwente.nl} \\
}

\begin{document}
\maketitle

\begin{abstract}
Digital neuromorphic processors are emerging as a promising computing substrate for low-power, always-on EdgeAI applications. In this tutorial paper, we outline the main architectural design principles behind fully digital neuromorphic processors and illustrate them using the SENECA platform as a running example. Starting from a flexible array of tiny RISC-V processing cores connected by a simple Network-on-Chip (NoC), we show how to progressively evolve the architecture: from a baseline event-driven implementation of fully connected networks, to versions with dedicated Neural Processing Elements (NPEs) and a loop controller that offloads fine-grained control from the general-purpose cores. Along the way, we discuss software and mapping techniques such as spike grouping, event-driven depth-first convolution for convolutional networks, and hard-attention style processing for high-resolution event-based vision. The focus is on architectural trade-offs, performance and energy bottlenecks, and on leveraging flexibility to incrementally add domain-specific acceleration. This paper assumes familiarity with basic neuromorphic concepts (spikes, event-driven computation, sparse activation) and deep neural network workloads. It does not present new experimental results; instead, it synthesizes and contextualizes findings previously reported in our SENECA publications to provide a coherent, step-by-step architectural perspective for students and practitioners who wish to design their own digital neuromorphic processors.
\end{abstract}




\section{Introduction}

Interpreting complex sensory data patterns in real-time and always-on with minimum power consumption is important for the survival of any living creature. The brain is responsible for performing this computation and has evolved over millions of years to be efficient in power consumption and processing speed. A honey bee's brain uses less than 1 milliwatt of power, yet it can perform a wide range of complex tasks such as navigation, communication, learning, and memory in real-time. The brain seamlessly integrates memory and learning and constantly evolves through experience. Despite the fact that ions, which are the signal carriers in the brain, move a million times slower than electrons, biological brains outperform electronic computers in terms of power consumption and latency. This document promotes the biological brain's architecture as a blueprint for future computing technologies, especially in the field of EdgeAI.

The brain is composed of a complex network of neurons that work together to create a highly efficient dataflow architecture. This architecture is designed for distributed processing and memory, enabling efficient parallel processing across a large three-dimensional structure. Neurons communicate through spikes, which allows for reliable digital data exchange to overcome the noisy environment. Moreover, this communication is sparse, which allows data to propagate and be processed quickly and efficiently. Although the brain is capable of executing any computational task, it specializes in processing natural signals such as audio and video.

\subsection{Promises of Neuromorphic engineering}

Neuromorphic engineering is a field that aims to enhance the speed and efficiency of computers by processing data intelligently. The main principle behind this approach is activation sparsity, which involves processing only the most significant data while avoiding energy consumption on ineffective operations. This approach is similar to how our brains work, and not only saves power but also enables these systems to respond quickly, making them ideal for fast-acting applications.

Neuromorphic processors are particularly useful for applications that deal with sparse data. Sparsity can be found in natural signals such as audio and video. In these signals, not every point in time holds important information. Similarly, the neurons in the brain are usually only 1 to 10\% active at any given time \cite{quian2010measuring}. 

To mimic this efficiency, neuromorphic processors are designed to be event-driven. They only process data when an event occurs, and in the absence of events, the processor idles. This feature allows them to consume power proportionally to the number of events they receive. This is known as event-driven power consumption.

The co-localization of memory and processing is a key feature that improves the efficiency of neuromorphic processors. This is accomplished by placing memory and processing units near each other, which is similar to how the human brain functions. This design significantly reduces the amount of data that needs to be moved, which is the primary factor that consumes power in most computing platforms \cite{indiveri2015memory}. 

Both sparsity exploitation and distributed parallel processing pose several challenges in developing algorithms, programming, and fabricating platforms. This is mainly because the mature processing ecosystem stack does not support such models efficiently. For instance, training Spiking Neural Networks in GPUs, the most accessible deep learning accelerators, is an extremely slow process. These challenges have hindered the growth and integration of neuromorphic systems into today's market \cite{christensen20222022}.

\subsection{Digital or Analog}

In this tutorial, we will be discussing the development of digital neuromorphic processors. While "neuromorphic processing" is commonly associated with "analog computing", it has been shown that integrating neuromorphic principles in the design of traditional digital processing systems can significantly improve their performance \cite{schuman2022opportunities}. Moreover, digital neuromorphic processors can pave the way for analog neuromorphic computing and can be scaled and developed in areas that are currently not feasible with analog computing.

Digital design offers several advantages over analog design. One of the most significant benefits is its compatibility with advanced technology nodes. As semiconductor manufacturing progresses, digital systems can be built using smaller, more efficient, and faster transistors. Digital neuromorphic systems can thus take full advantage of the latest manufacturing techniques, which can enhance their performance and reduce costs. In contrast, analog designs require a complete redesign to adapt to each new technology node, and as the size of analog components decreases, they become more susceptible to noise, which can significantly affect their performance \cite{furber2023digital}.

Digital logic blocks can be easily duplicated and reused in different designs, which simplifies the process of creating complex architectures. This means that once a digital circuit is designed for a neuromorphic function, it can be effortlessly integrated into multiple systems. Consequently, digital neuromorphic processors are highly reconfigurable, making them more adaptable than analog systems and better equipped to support a wide range of applications.

The main benefit of using analog neuromorphic designs is their low power consumption. Analog circuits can operate with extremely low power, which is especially useful when energy conservation is a priority. However, when it comes to overall performance and energy efficiency, digital systems may actually be more effective \cite{sun2023analog}. This is because digital systems can utilize superior technology nodes and employ more complex and precise designs. Combining digital and analog systems in a single neuromorphic processor can sometimes provide the best of both worlds, exploiting the power efficiency of analog circuits with the precision and scalability of digital design \cite{le202364}. However, choosing between digital or analog implementations should be based on the specific needs of the application, as each approach has unique contributions to the development of neuromorphic engineering.

The field of digital neuromorphic chips is rapidly advancing, with various companies and research groups developing successive generations of technology. SpiNNaker\cite{furber2014spinnaker}, a chip developed as part of the Human Brain Project, now has an updated version called SpiNNaker2\cite{mayr2019spinnaker}. Intel's neuromorphic chip, Loihi\cite{davies2018loihi}, has also seen an upgrade in the form of Loihi2\cite{orchard2021efficient}, indicating significant progress in digital neuromorphic technology. IBM has made significant contributions to the development of neuromorphic computing through its TrueNorth chip \cite{akopyan2015truenorth}. The company has introduced a successor to the TrueNorth chip named NorthPole \cite{modha2023neural}. IMEC's initial venture into this space with the uBrain processor \cite{stuijt2021mubrain} has led to the development of a more advanced chip called SENECA \cite{yousefzadeh2022seneca}.

The field of digital neuromorphic processing technology is being revolutionized by a number of startups that are introducing EdgeAI solutions to the market. Among these startups is GrAI Matter Labs, which was acquired by Snap. They have developed the NeuronFlow architecture \cite{moreira2020neuronflow}, followed by GrAI-VIP as a second generation. Synsense, another neuromorphic startup, initially designed analog neuromorphic processors and now also offers digital neuromorphic processors \cite{richter2023speck}. BrainChip is continuously scaling and improving its AKIDA architecture, which is a fully digital neuromorphic processor \cite{demler2019brainchip}. These examples illustrate a trend of continuous development in digital neuromorphic chips, with each new release aiming to provide better performance and energy efficiency.


\subsection{The SENECA Project}

This document is a compilation of the findings of the SENECA project, an ongoing initiative in the field of neuromorphic engineering that has been in progress for several years. It presents the results from previously published papers, which reflect the continuous efforts and progress made by the project's team of researchers and university collaborators \cite{yousefzadeh2022seneca, molendijk2022benchmarking, yousefzadeh2022energy, tang2023seneca, tang2023open, kevin_paper}. 

The SENECA project began in the summer of 2020 and has so far been funded by seven research grants from the European Union (€1.7M), The Netherlands Enterprise Agency (€0.8M), and IMEC (€1.5M): TEMPO\cite{TEMPO}, ANDANTE\cite{ANDANTE}, MNEMOSENE\cite{MNEMOSENE}, DAIS\cite{DAIS}, MeM-Scales\cite{MeM-Scales}, REBECCA\cite{REBECCA}, NEUROKIT2E\cite{NEUROKIT2E} and NimbleAI\cite{NimbleAI}. Although it is still in progress, the project has already made significant contributions to the field, and further advancements and insights are expected in the future.

\section{The main architectural template: Array of Tiny Processors}

\begin{figure}
    \centering
    \includegraphics[width=0.8\linewidth]{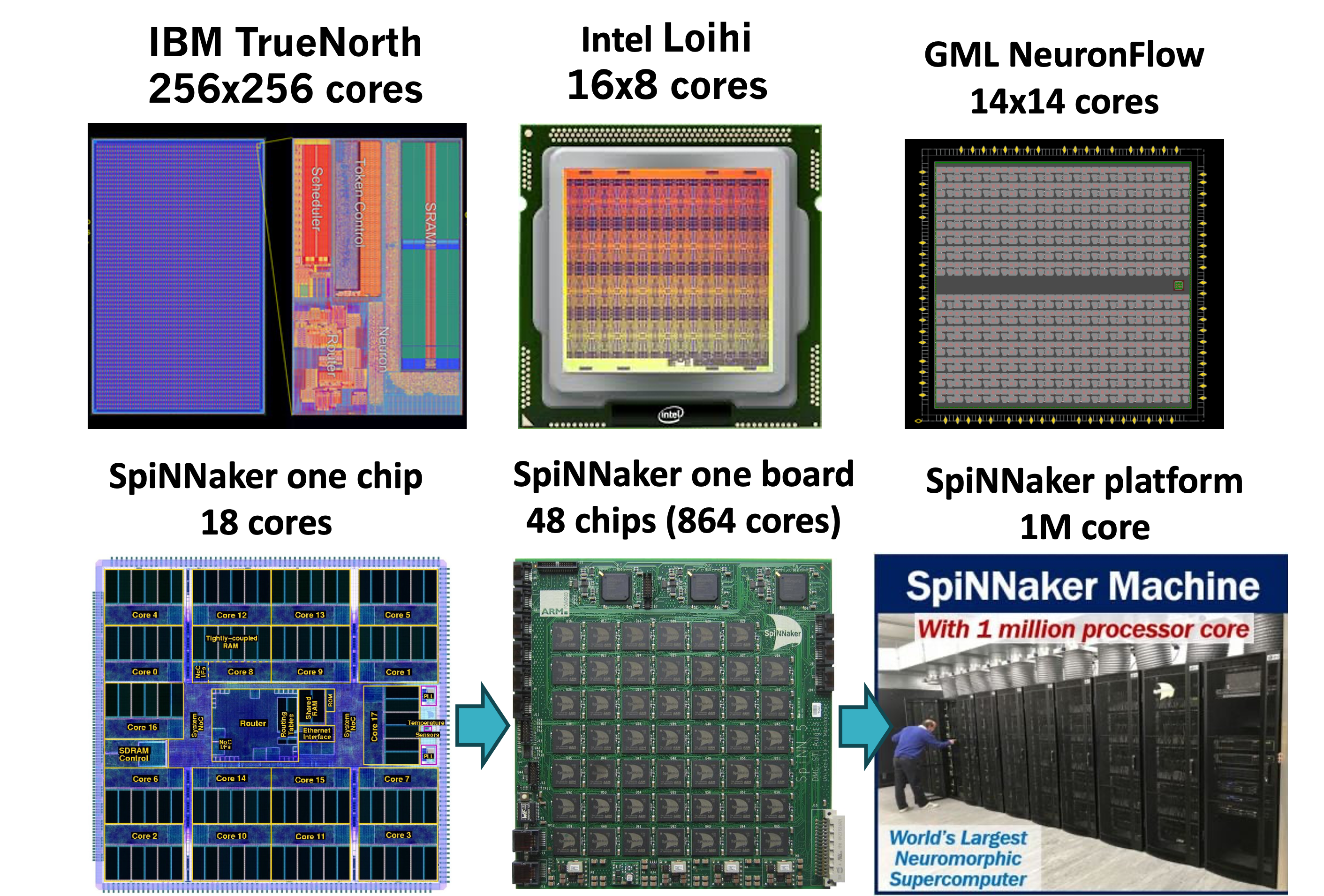}
    \caption{Examples of several digital neuromorphic chips. All follow the same template of connecting many tiny processing elements to build a scalable distributed processing platform.}
    \label{fig:tn_lh_spn}
\end{figure}

Advanced digital neuromorphic processors are typically composed of a network-on-chip that interconnects many small processing elements. The image in Fig.\ref{fig:tn_lh_spn} displays several digital neuromorphic processors that were designed using this same configuration. This structure emulates the decentralized architecture of the biological brain, resulting in distributed memory and processing power across the processor. In neuromorphic platforms, each core communicates with others by sending and receiving data packets called spikes or events. Thus, to create neuromorphic processors, two essential components are required: the tiny processing core and the Network on Chip (NoC). The processing cores function as the fundamental units of computation, similar to neurons in the human brain, while the NoC serves as the communication backbone, analogous to the synaptic connections within the brain's network.

\subsubsection{Tiny processor core}

Neuromorphic processors have independent and asynchronous cores that can process input events and generate output events in parallel\footnote{Here, asynchrony refers to a processing model where individual threads execute on each processing core without explicit synchronization barriers. Neuromorphic processors may have synchronized circuits but retain an asynchronous execution model.}. Each core has its own memory and processing logic, and some may also have control units that manage the overall process.

The functionality of a platform is mainly determined by the design of its neuromorphic processing core. These functionalities include the neuron models and activation functions (e.g., LIF, ReLU, LSTM, and GRU), the neural network architectures (e.g., Dense, Conv, Recurrent, and Transformer), the number of neurons and weights, as well as the precision of these weights. The architecture of a processing core has a significant impact on the performance, area, and energy consumption of the system. Designing such a core involves various trade-offs, which are listed in \cite{tang2023seneca} and will be briefly discussed here. 

A neuromorphic processing core can emulate several neurons in a neural network by \textbf{time-multiplexing} their fast data path. This is possible because a single processing core can work much faster than a biological neuron. Though time-multiplexing is not biologically plausible, it is commonly used in the design of neuromorphic cores, as it improves the area efficiency of the system. Designers have the option to increase the time-multiplexing ratio to enhance area efficiency and reduce communication overhead, or decrease it to improve performance and energy efficiency. This trade-off is an essential consideration for designers.

When developing a processing core, there is a crucial decision to be made between \textbf{flexibility and performance}. A core that prioritizes extremely high performance may be limited to a narrower range of applications and cannot benefit from future software optimizations. Therefore, this decision isn't always straightforward and is often misunderstood. This is because it is possible to improve system performance by leveraging the flexibility of an architecture. Several recent works demonstrate superior performance achieved by deploying neural networks on commercial-off-the-shelf CPUs/GPUs, compared to domain-specific deep learning accelerators or neuromorphic processors (e.g., \cite{NeuralMagic}, \cite{knight2018gpus}, \cite{chen2020slide}, \cite{shukla2019remodel}). 

In this tutorial, we start with a highly flexible neuromorphic core. We then integrate domain-specific accelerators to enhance performance for specific benchmarks, such as fully connected and convolutional layers. This approach allows us to continually enhance the performance of our neuromorphic platform and keep up with the dynamic changes in the neuromorphic algorithm domain. With this methodology, the core can meet the high-performance requirements of target algorithms while remaining flexible enough to deploy new algorithms or software optimizations. However, the drawback is that it may require more silicon area compared to fully dedicated cores.

SpiNNaker is an excellent example of a neuromorphic platform that adopted such an approach. The initial version of the chip utilized tiny and low-power ARM processors as the neuromorphic processing cores. As the next step, SpiNNaker2 has been developed with several new accelerators added to the ARM processors to improve its performance for specific applications while maintaining its general applicability for new applications. SpiNNaker proved to be a valuable endeavor, as the scaled-up platform is versatile enough to handle all kinds of neuromorphic computations, including various neuron models \cite{yousefzadeh2018performance}, neural networks \cite{kelber2020mapping}, and learning algorithms \cite{rostami2022prop}.

\begin{figure}
    \centering
    \includegraphics[width=0.5\linewidth]{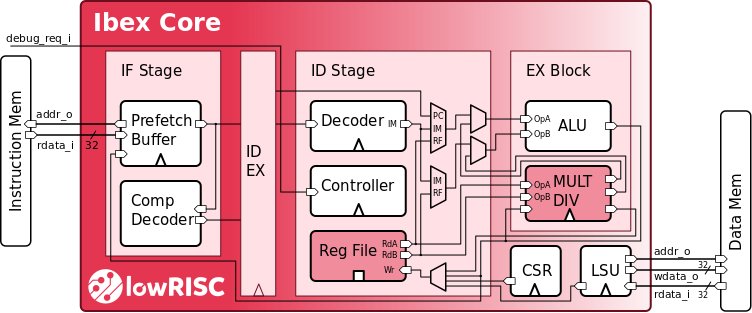}
    \caption{Internal pipeline and architecture of IBEX, from \url{iis-projects.ee.ethz.ch}.}
    \label{fig:ibex}
\end{figure}

As the core of the SENECA project, we used tiny RISC-V processors as our processing cores. Our main goal was to quickly build a flexible neuromorphic platform capable of running various applications. After conducting some benchmarking on the Epiphany platform (which is also built by connecting an array of RISC processors with NoC) and the SpiNNaker platform (which uses ARM processors), we decided to select the IBEX core (Fig.\ref{fig:ibex}). It demonstrated a reasonable area/performance trade-off and had excellent open-source repositories.

\begin{figure}
    \centering
    \includegraphics[width=1\linewidth]{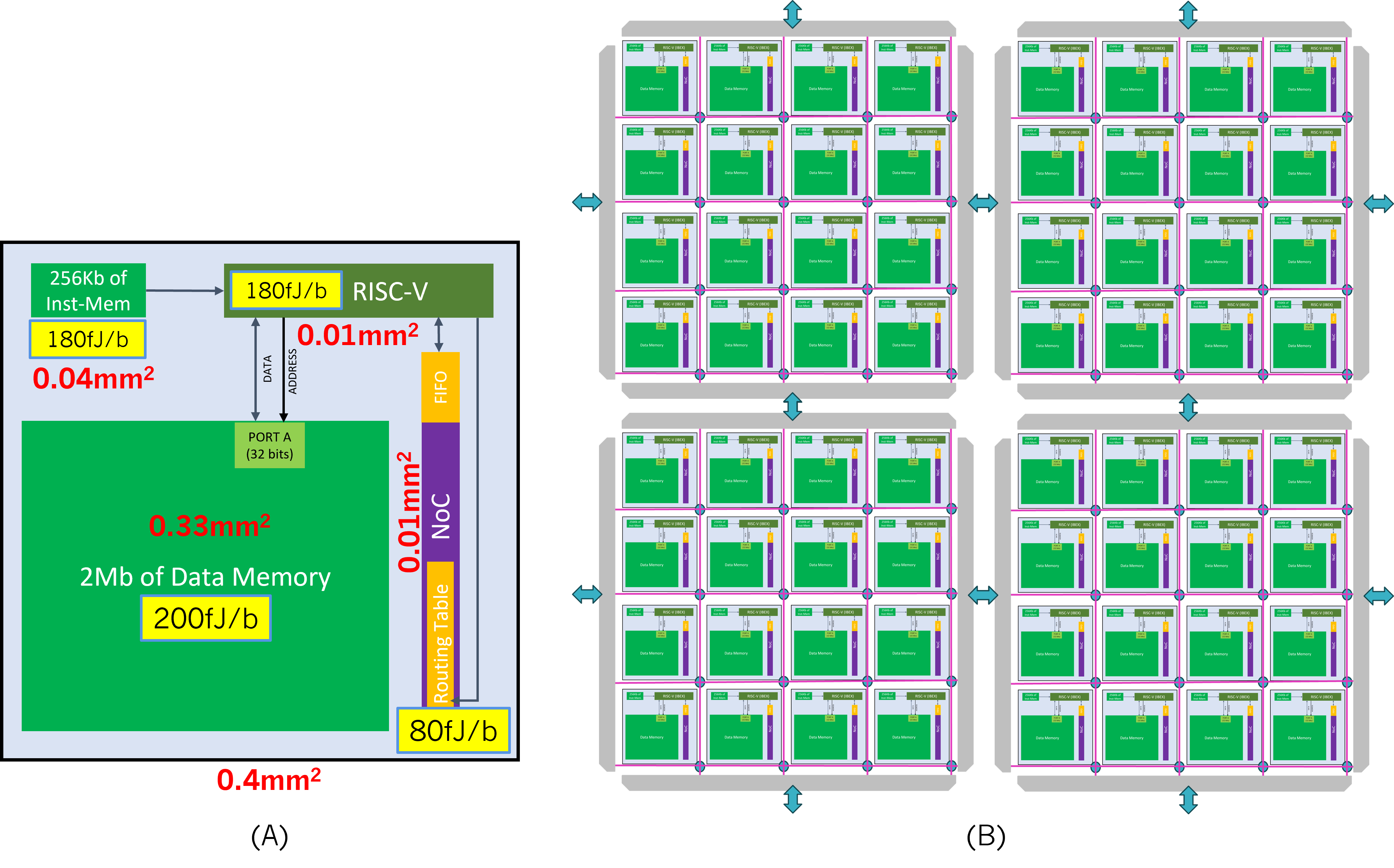}
    \caption{On the left, you can see the initial version of our neuromorphic processing core, while on the right, you can see the complete platform created by linking the cores together. The core architecture is annotated with the area and energy figures in the GF-22nm FDSOI technology node.}
    \label{fig:array_of_rv}
\end{figure}

The simplified architecture of the proposed neuromorphic processor is illustrated in Fig. \ref{fig:array_of_rv}. It consists of several interconnected processing cores, each of which includes RISC-V and NoC (will go through it in the next subsection), as well as instruction and data memory. The execution model and neural network mapping are stored in the instruction memory, while the data memory stores the synaptic weights and neuron states. As planned, we began with the most basic and adaptable neuromorphic processor we could conceive.

\subsubsection{Network on Chip}

The Network on Chip, also known as NoC, is a technology that allows for efficient, flexible, and scalable connections between neuromorphic processing cores. In biological brains, processing cores can connect and disconnect during operation through flexible plastic wires. However, in electronic circuits, it is not possible to form completely new connections after fabrication. Therefore, all cores must be connected through wires by default. These wires are time-multiplexed among cores since the electrical communication speed is much faster than ions in the brain. NoC handles this time multiplexing to ensure an effective exchange of information between processing cores.

In neuromorphic platforms, the neurons are connected via shared wires. As a result, the generated spikes cannot be a single electrical pulse. Instead, each spike leaving the core must carry its unique source neuron ID to ensure proper decoding in the destination core. This means that each spike traveling through the NoC (Network on Chip) is actually a packet of data rather than a single pulse. This data packet is commonly known as AER (Address Event Representation) in many neuromorphic processors \cite{yousefzadeh2017multiple}.

The main purpose of the Network on Chip (NoC) is to ensure that data (spike) is delivered promptly to enable the processing core to perform its functions as quickly as possible. After observing various neuromorphic platforms such as NeuronFlow, Loihi, SpiNNaker, and Epiphany, we found that the NoC has never been a performance bottleneck in those platforms. The high level of sparsity in our algorithms and the high operation density\footnote{Operation density is defined as the number of operations required to process a data packet.} for each spike could be the reason for this. 

In SENECA, we observed that the Processing core (RISC-V) shown in Fig.\ref{fig:array_of_rv} was always our primary bottleneck in terms of both performance and energy consumption. We did not find that NoC significantly contributed to power consumption or performance. This is because when the operation density is high, transferring a data packet consumes much less energy/time than processing it. Based on these observations, we believe that any simple NoC that can flexibly connect the neuromorphic cores would be sufficient for our purpose. For instance, LOIHI \cite{davies2018loihi}, TrueNorth\cite{akopyan2015truenorth}, and NeuronFlow \cite{moreira2020neuronflow} use a simple packet-switched mesh type NoC. There are a number of open-source NoCs that can be used to construct a neuromorphic platform directly \cite{RaveNoC, papamichael2015connect, AwesomeMesh}.

NoC, despite not affecting performance and energy consumption in the system, can have various side effects on the performance of other system elements. For instance, it can create a high memory overhead to store the routing table. Each neuron in the core requires its set of destination addresses, defined by the network architecture and mapping. For instance, the TrueNorth \cite{akopyan2015truenorth} uses 26 bits for each neuron to designate a single destination core out of a possible 64 million cores in the system. This approach not only requires a considerable amount of memory but also lacks flexibility since each neuron can only be assigned to a single destination core. Another example is NeuronFlow \cite{moreira2020neuronflow}. Despite the flexible routing and axon-sharing mechanism in NeuronFlow, the NoC routing table still occupies about 25\% of the total memory.

SpiNNaker \cite{furber2014spinnaker} uses a routing method called source-based addressing. Instead of the core knowing the destination address, each NoC has a routing table with enough information to forward the packet to its correct output ports. Therefore, the information about the final destination addresses of each neuron is distributed over many routers and can be optimized to reduce data movement and the size of routing tables. As a result, the routing information required for a structured neural network architecture (like CNN and Dense neural networks) can be much smaller.

A source-based addressing NoC has a second advantage of allowing efficient multicasting of the spike when the spike has multiple destination cores. Without multicasting, the core that generated the spikes would need to send multiple copies of the packet from the source to the destination, which could lead to network congestion. In the source-based addressing scheme, the packet can be copied to multiple output ports of the nearest router to reach the destinations efficiently. However, the improvement provided by this multicasting is negligible when the NoC is not congested.

\begin{figure}
    \centering
    \includegraphics[width=1\linewidth]{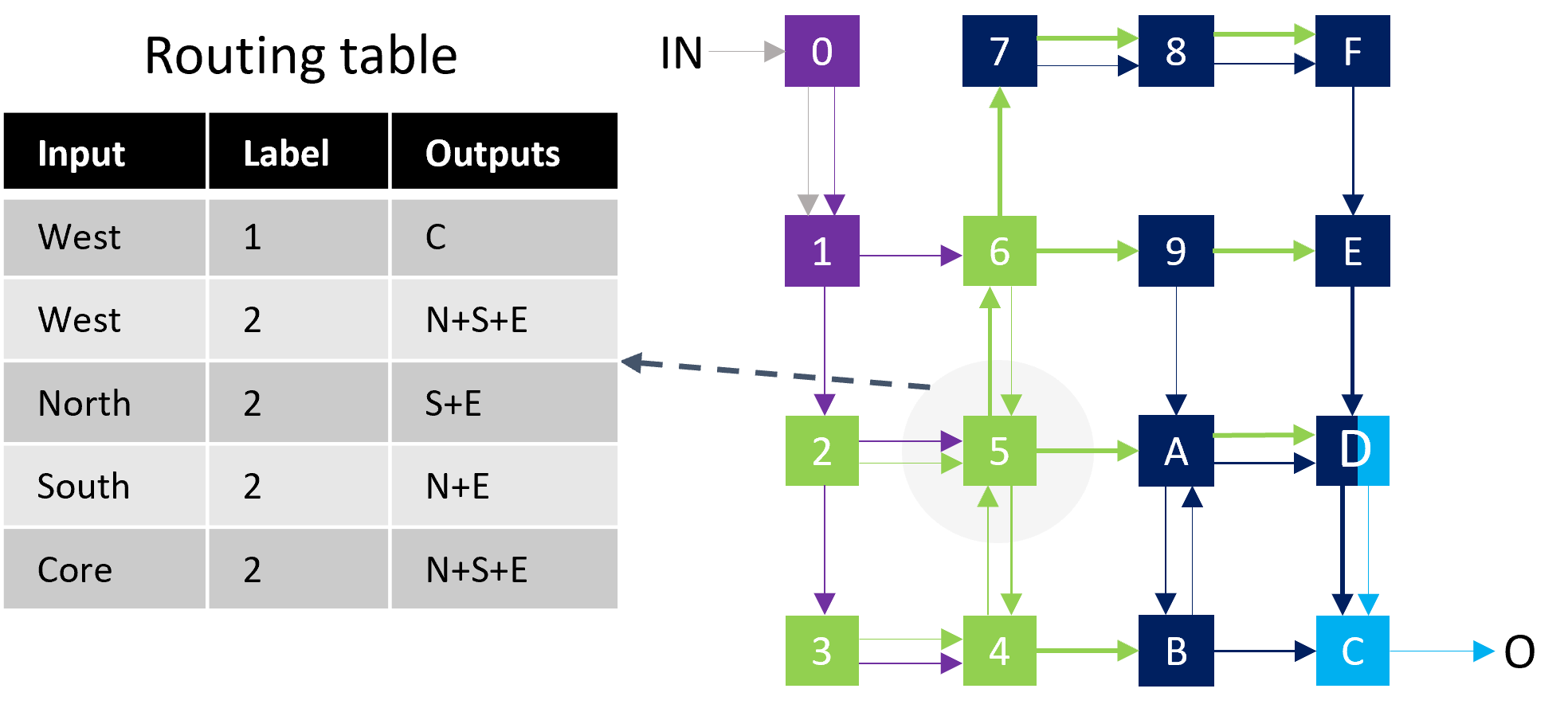}
    \caption{On the right, you can see a 4-layer neural network being mapped on 16 cores (color-coded). On the left, there is a routing table of the NoC in the core 5. The NoC processes the incoming packet based on the input port and the label it carries. It then routes them to one or several output ports, depending on the content of the routing table. The label is a part of the spike packet.}
    \label{fig:NoC_mapping}
\end{figure}

SENECA has implemented a similar source-based addressing scheme as SpiNNaker, but with a simpler NoC design. We opted for a mesh scheme with one NoC per core. The SENECA NoC concept is illustrated in Fig. \ref{fig:NoC_mapping}, which features a small routing table per NoC. Consequently, each spike packet must carry a label as its source address. Here, the label represents the source layer ID as shown in Fig. \ref{fig:NoC_mapping}.

\section{Event-driven processing in our neuromorphic platform}
\begin{figure}
    \centering
    \includegraphics[width=0.3\linewidth]{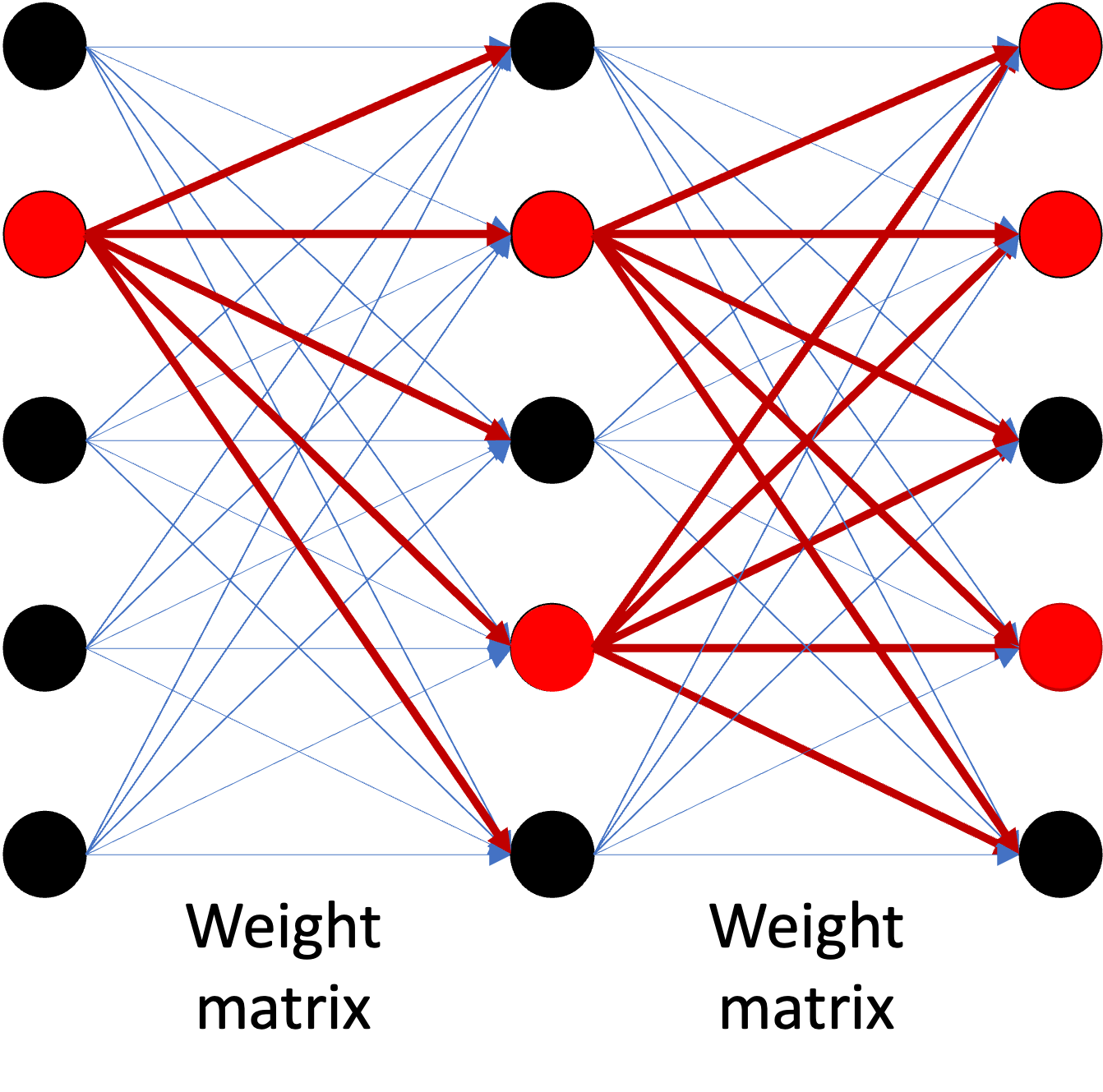}
    \caption{Event-driven processing of a fully connected layer. Neurons with non-zero activations that generate a spike (event) are shown in red. These events move to the next layer and update the corresponding receiving neurons.}
    \label{fig:fc_layers}
\end{figure}

We now have all the necessary components to build our neuromorphic platform, including tiny processing cores and a network-on-chip. We can proceed to implement a simple event-driven, fully connected neural network and measure its performance. Although fully connected neural networks are basic, they serve as a good representation of neural network accelerators' performance, as they form the foundation of any neural network. These networks require vector-matrix multiplication, where the vector represents the activation output of the previous layer and the matrix represents the synaptic weights that connect the previous layer to the next layer (as shown in Fig. \ref{fig:fc_layers}).

In the computation of vector-matrix multiplication using an event-driven approach, we first break down the activation vector into its non-zero elements. Then, we perform several scalar-vector multiplications where the vector represents the weights of synapses that connect this activation to the next layer (these synapses are highlighted in bold red in Fig. \ref{fig:fc_layers}). Each non-zero activation forms a spike packet that reaches the destination neurons and updates them based on the synaptic weight. These spike packets can be binary (like their biological counterparts) or can have values. In the case of binary spikes, the amount of synaptic weight will be added to the neuron state of the next layer (known as the partial sum of the neuron in DNN terminology). However, if the spike packet contains a value, this value needs to be multiplied by the synaptic weight before being added to the neuron state of the next layer.

There is an ongoing debate regarding the use of spikes with value, commonly known as valued or graded spikes. These types of spikes are controversial because they are not directly biologically plausible. However, using valued spikes can increase the accuracy of neural networks. This improvement comes at the cost of sending more bits over the NoC and performing a multiplication operation. Despite this drawback, valued spikes can potentially enable neural networks to achieve the same level of performance as binary spikes while using fewer spikes overall. This is why the second generations of Loihi, TrueNorth, and SpiNNaker all allow the use of graded spikes. In our case, since we are using a RISC-V, both options are available to us.

\begin{table}
    \centering
    \begin{tabular}{|c|c|c|}
        \hline
         \textbf{Processor} &  \textbf{Inference time ($\mu s$)} & \textbf{Energy ($\mu J$)} \\
         \hline
         Loihi1     &  3378               & 372          \\
         \hline
         SpiNNaker2 &  1000               & 7.1          \\
         \hline
         \textbf{SENECA(V1)} &  \textbf{7000}               & \textbf{34}           \\
         \hline
    \end{tabular}
    \caption{Comparison of SENECA version 1 (RISC-V and NoC) with other neuromorphic platforms using a fully connected neural network application. More details about this experiment are available in \cite{kevin_paper}.}
    \label{tab:benchmark_first}
\end{table}

As a first step in the SENECA project, we used a simple benchmarking task that was previously deployed and benchmarked using Loihi1 \cite{blouw2019benchmarking} and SpiNNaker2 \cite{yan2021comparing}. This activity allowed us to measure the performance of this simple neuromorphic chip in comparison to other, more advanced versions. The results are reported in \cite{kevin_paper} and can be found in Table \ref{tab:benchmark_first}. Please note that benchmarking on Loihi2 is not available.

In comparison to Loihi, our neuromorphic platform is slower by a factor of 2. This is because Loihi has a custom-made processing core with specialized instructions for the neural network. On the other hand, RISC-V requires many execution cycles to finish one neuron update. However, surprisingly, the energy consumption of SENECA(V1) is much better than that of Loihi. We believe the main reason is that Loihi's processing cores emulate a complex neuron model along with learning capabilities, which were not utilized in this benchmark. This fact indicates that a more dedicated (less flexible) processing core can only result in good performance if there is a high degree of matching between the application and the hardware. In Loihi2, the neuron model is programmable to increase system flexibility.

When compared to SpiNNaker2, our SENECA(V1) platform consumes significantly more energy and takes longer to complete inference. The primary reason for this difference is that SpiNNaker2 has a dedicated neural processing accelerator that can update neurons much faster than a general-purpose processor. This observation has provided us with valuable insights into how we can optimize our neuromorphic platform.

\section{Adding Hardware Accelerators}

Although our neuromorphic platform is already complete with RISC-V processing cores and NoC, we want to demonstrate how adding dedicated accelerators can improve the platform's performance without sacrificing its programmability. Adding accelerators is beneficial if there's extra space in the silicon area, as they boost the system's performance when used and can be turned off when not in use.

\subsection{Neuron Processing Elements}

\begin{figure}
    \centering
    \includegraphics[width=0.5\linewidth]{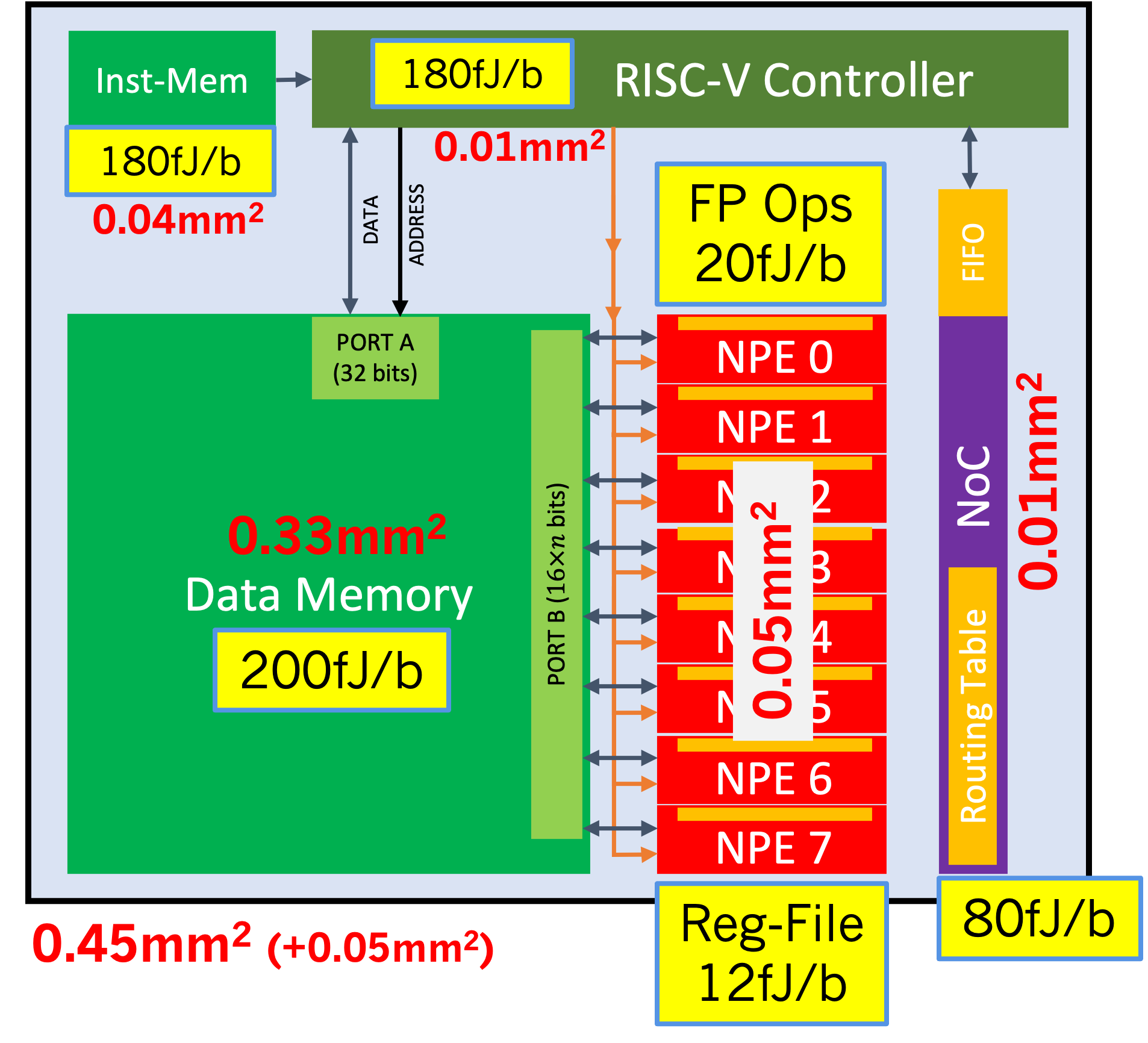}
    \caption{Neural Processing Elements (NPEs) have been integrated into our neuromorphic processors to build SENECA(V2). The figure now displays the energy and area consumption of the processing core after adding the NPEs. Additional details about the benchmarking results can be found in \cite{tang2023open}.}
    \label{fig:adding_NPEs}
\end{figure}

In SENECA, the first hardware accelerator added was a set of Neural Processing Elements (NPEs). These NPEs work synchronously, executing the same instructions on various data to perform efficient vector operations. Fig. \ref{fig:adding_NPEs} shows the architecture of the processing core, with 8 NPEs added to it. The figure is also annotated with the energy and area consumption \cite{tang2023open}. However, NPEs are not a complete processor and require a controller to fetch and decode operations for them. Therefore, in SENECA(V2), RISC-V serves as the controller in the system and does not perform the neural computations anymore.

NPEs (Neural Processing Elements) execute Brain-Float (16b) operations by default, even though parameters can be stored in lower resolution. Using a 16-bit integer for neuron states (partial sums) was insufficient due to the limited range. BrainFloat16 operations consume more power than integer operations but provide a significantly higher range to avoid overflowing of neuron states. Floating point operations provide a mixture of linear and logarithmic quantization schemes that can be optimally tuned to obtain the best accuracy/performance trade-off. As shown in Fig. \ref{fig:adding_NPEs}, the floating point operations in NPEs still consume 10x less power compared to data memory access. Due to the nature of being dedicated compute elements, they also consume much less than executing INT32 operations in RISC-V. You can find the full instruction set and measurements in \cite{tang2023seneca}.

\begin{table}
    \centering
    \begin{tabular}{|c|c|c|}
        \hline
         \textbf{Processor} &  \textbf{Inference time ($\mu s$)} & \textbf{Energy ($\mu J$)} \\
         \hline
         Loihi1     &  3378               & 372          \\
         \hline
         SpiNNaker2 &  1000               & 7.1          \\
         \hline
         SENECA(V1) &  7000               & 34           \\
         \hline
         \textbf{SENECA(V2)} &  \textbf{1100} & \textbf{7} \\
         \hline
    \end{tabular}
    \caption{Deployment of the benchmarking application of Table \ref{tab:benchmark_first} on SENECA(V2)}
    \label{tab:benchmark_NPEs}
\end{table}

The eight parallel NPEs (Neural Processing Elements) can perform eight complex neural operations simultaneously, which is expected to significantly enhance the system's performance. Furthermore, using BF16 instead of INT32 and the ability of a single controller to handle all eight NPEs significantly reduced the overall power consumption. The benchmarking results when using the NPEs in the SENECA(V2) platform are presented in Table \ref{tab:benchmark_NPEs}. As shown, we have achieved almost the same performance as SpiNNaker2. This is not surprising, as the main difference between SENECA(V1) and SpiNNaker2 was the existence of the vector processing unit in SpiNNaker2. In \cite{kevin_paper}, it has been demonstrated that the averaged energy consumption of RISC-V for one inference has reduced from 30$\mu s$ to 3$\mu s$, while NPEs only added 2us to execute all the neural updates.  

The NPEs used in the SENECA project can be improved. One issue is the long pipeline (4 stages) caused by the use of floating-point ALU, resulting in inefficiency when there is a pipeline hazard. Another area for improvement is data conversion. Although the NPEs support parameters and operations in lower resolution (integer and flex point), converting between formats adds significant overhead. The final challenge is related to the compilation of NPE instructions. While RISC-V has its own advanced computer that could be used for this project, NPEs with their own instruction sets require their own compiler if the user wants to write code in higher-level languages. Currently, we are using assembly code for NPEs in the SENECA project, which slows down the development of new applications.

\subsection{Loop Controller}

After we added the Processing elements, we noticed that RISC-V utilized 3$\mu J$ of energy, while NPEs only used 2$\mu J$ for all the neural operations. Even though RISC-V is more flexible, executing instructions in it is quite expensive. So, we decided to analyze the C program of RISC-V to gain greater insight. During profiling, we discovered that RISC-V provides a small set of instructions to NPEs in a nested loop. Although the instructions were very similar, each instruction needed to be individually loaded from instruction memory, fetched and decoded, which added a lot of overhead to the system. 

To improve the efficiency of executing loops in computer architecture, a solution known as a loop buffer has been traditionally used. Inspired by this, we developed a new independent controller, which we named the loop controller. This controller is specifically designed to handle neural network loops and has its own small instruction memory constructed using a register file that consumes less power than SRAM. We later referred to this concept as a hierarchical control system \cite{tang2023seneca}, where the RISC-V controls the Loop controller, which in turn controls the NPEs.

\begin{figure}
    \centering
    \includegraphics[width=1\linewidth]{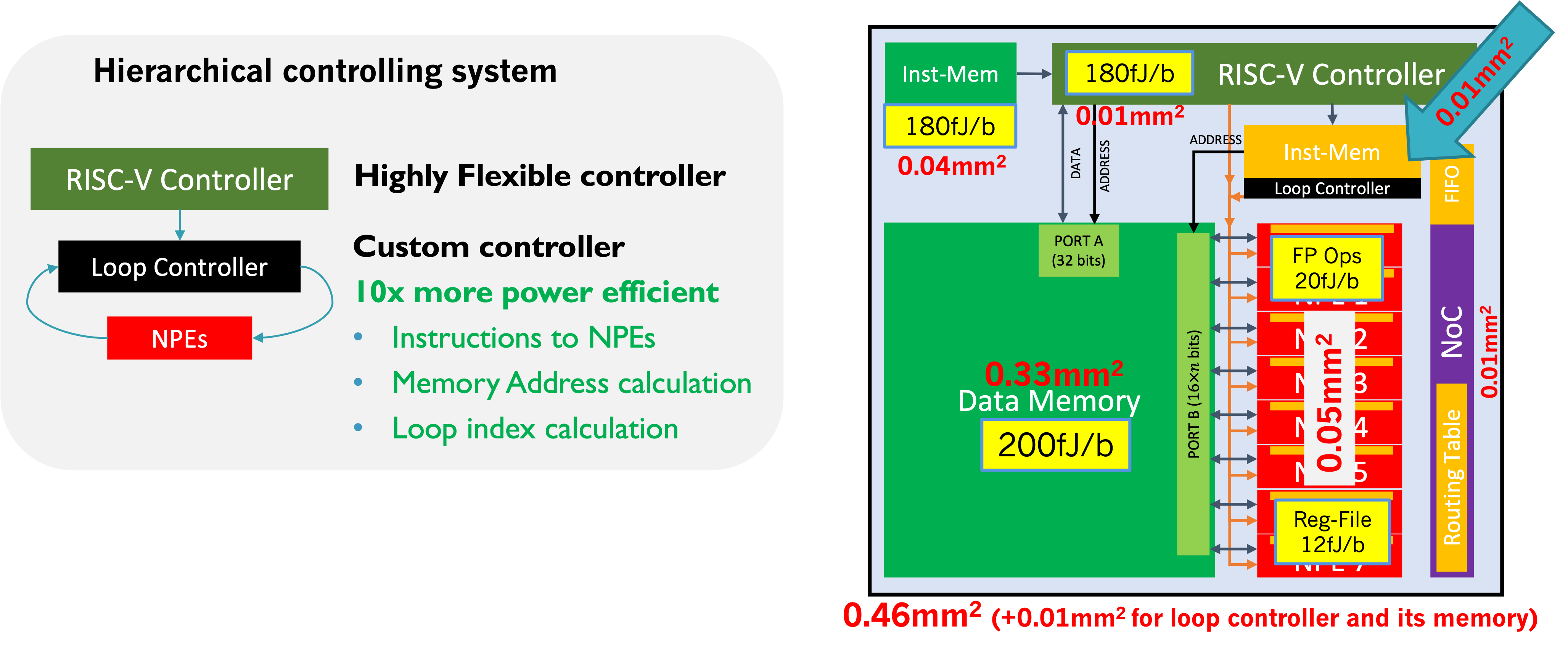}
    \caption{Adding loop controller to SENECA to make a new version: SENECA(V3)}
    \label{fig:loop_controller}
\end{figure}

Fig.\ref{fig:loop_controller} illustrates the SENECA(V3) architecture with the addition of a loop controller and its own instruction memory. The loop controller is highly power-efficient, being 10 times more efficient than RISC-V when executing loops and providing instructions to NPEs. It is designed to handle a limited set of instructions, specifically for managing loop indexes, nested loops, and memory address calculations.  

\begin{table}
    \centering
    \begin{tabular}{|c|c|c|}
        \hline
         \textbf{Processor} &  \textbf{Inference time ($\mu s$)} & \textbf{Energy ($\mu J$)} \\
         \hline
         Loihi1     &  3378               & 372          \\
         \hline
         SpiNNaker2 &  1000               & 7.1          \\
         \hline
         SENECA(V1) &  7000               & 34           \\
         \hline
         SENECA(V2) &  1100               & 7             \\
         \hline
         \textbf{SENECA(V3)} &  \textbf{550}  & \textbf{3}   \\
         \hline
    \end{tabular}
    \caption{Benchmarking result for SENECA(V3) which includes loop controller.}
    \label{tab:benchmark_loopcontroller}
\end{table}

Table \ref{tab:benchmark_loopcontroller} displays the performance improvement of SENECA(V3) when using the loop controller, as reported in \cite{kevin_paper}. The loop controller reduces the load on RISC-V, resulting in a decrease in energy consumption from 3$\mu J$ to 0.2$\mu J$.

After implementing the loop controller, we added several other accelerators to our system to boost its performance. You can find more details about these improvements in \cite{tang2023seneca}. As explained in the article, these accelerators improved the system's performance in neural network inference, which was our target benchmark. However, they did not compromise the flexibility of our SENECA(V1) platform. Therefore, we can follow the same incremental approach to enhance the performance of our platform for any other benchmarking applications. For instance, we already reported our initial results for on-device learning (using e-Prob) in \cite{tang2023seneca}, and the SENECA team is currently working on designing hardware accelerators to enhance the performance of on-device learning further.

It is possible to perform pre- and post-processing of applications directly on the RISC-V. Depending on the expected performance of the system, it may or may not be necessary to design an accelerator for these glue algorithms. An example illustrating this point will be discussed in the next section.

\section{Software optimizations}

One of the key advantages of the flexible processing core is its ability to enhance system performance by implementing advanced algorithmic and software optimizations. In addition to typical hardware-aware algorithm optimizations such as quantization and sparsification, this section will outline three other techniques that we utilized to improve system performance.

\subsection{Spike Grouping}

\begin{figure}
    \centering
    \includegraphics[width=0.7\linewidth]{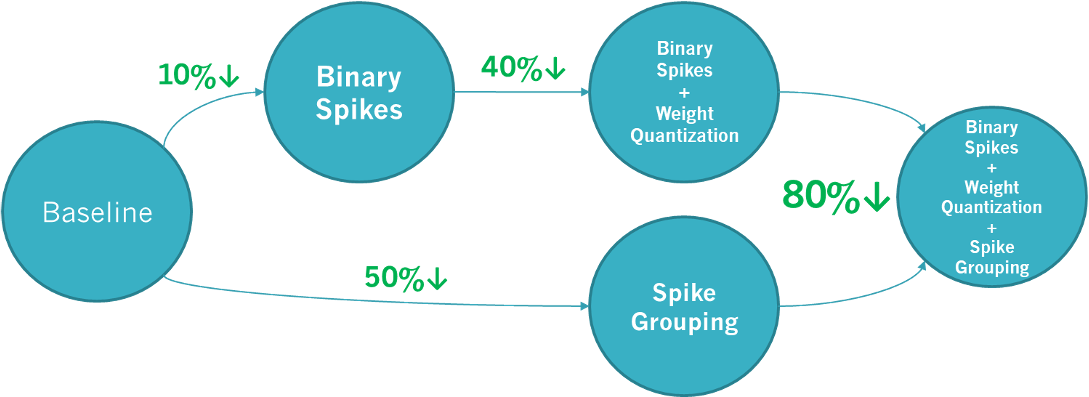}
    \caption{Performance (both energy and latency) improvement by using various optimization techniques. By using spike-grouping alone, the performance of SENECA(V3) can be improved by a factor of two in both energy and time. Additionally, using binary spikes can reduce power consumption by 10\%, while weight quantization (to 4 bits) can reduce power consumption by 40\%.}
    \label{fig:sw_opt_tech}
\end{figure}

Neuromorphic systems consume a lot of energy during memory access. When executing neural networks, it is common for the same data to be read and written multiple times in loops. To reduce memory access, GPUs use batch processing, reusing data as much as possible before discarding it. Similarly, in SENECA, a technique called spike-grouping is used to process several corresponding spikes at once, further reducing memory access. 

In Fig.\ref{fig:fc_layers}, we can observe that layer-2 produces two spikes which update all the neurons in layer-3. To process each spike, the neuron state of the neurons in layer 3 needs to be read, updated and written back. If we process both spikes simultaneously, the neurons in layer-3 will be read only once, updated twice and written back only once. Therefore, grouping two spikes reduces the memory access by reading and writing the neuron states only once for both spikes. We tried to group four spikes together (whenever possible), which resulted in a 50\% reduction in both energy consumption and inference time, as demonstrated in Fig.\ref{fig:sw_opt_tech}.

\subsection{Depth-First inference}


\begin{figure}
    \centering
    \includegraphics[width=0.99\linewidth]{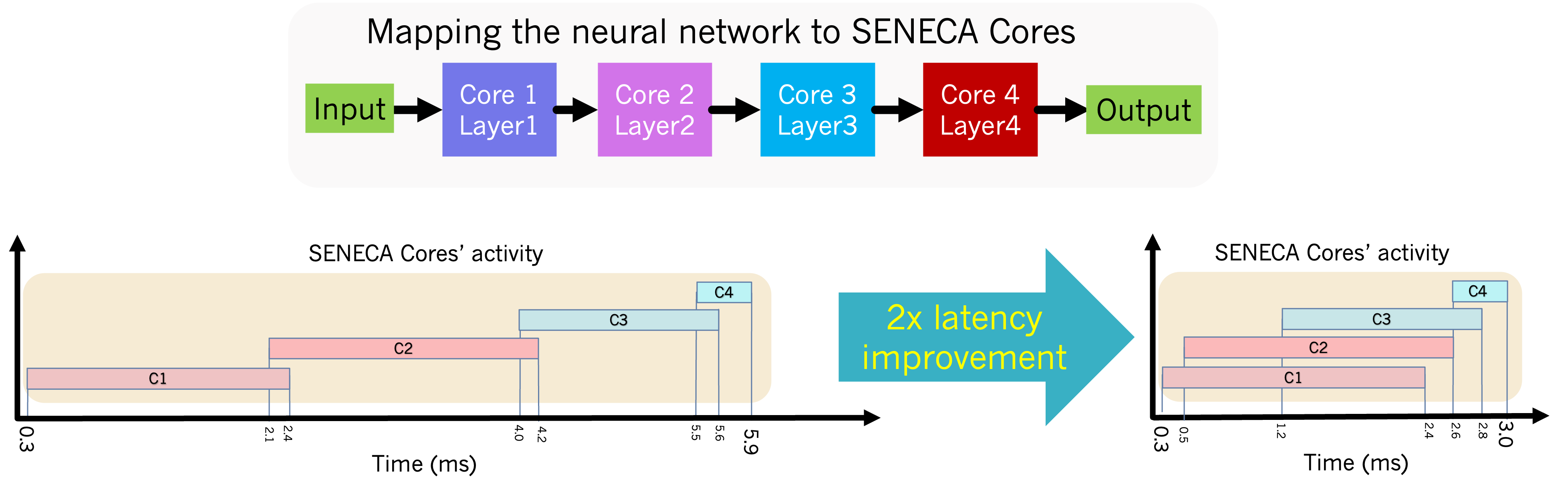}
    \caption{Using the event-driven depth-first method to map a 4-layer CNN on SENECA can reduce memory consumption and latency.}
    \label{fig:depth_first}
\end{figure}

Neuromorphic platforms face a problem with Convolutional Neural Networks (CNNs) that results in excessive memory usage for neuron states. CNNs are popular because they can use a few individual parameters for all synaptic weights in the network, reducing the number of parameters that need to be stored compared to the number of synapses. As a result, CNNs typically have a much larger number of neurons than the number of individual parameters. However, when CNNs are used in neuromorphic platforms, they often consume significantly more memory (e.g., 200 times) compared to their deployment in a DNN accelerator. This is because neuromorphic processors allocate dedicated memory to each individual neuron.

Despite their memory limitations, Convolutional Neural Networks (CNNs) are crucial in the field of neuromorphic computing when it comes to processing visual data. However, the SENECA platform offers flexibility in implementing hybrid networks where some neurons do not have dedicated memory. An event-driven depth-first inference was developed to avoid using dedicated memory for each neuron in the feedforward CNN layers. This allows for efficient deployment of CNNs on multi-core mapping, resulting in a significant reduction in both latency and memory usage.

The Event-driven Depth-first method is explained in detail in \cite{kevin_paper}. Fig.\ref{fig:depth_first} (top) illustrates how four CNN layers are mapped onto four SENECA cores (one layer per core). The depth-first inference fuses the operations of layers, allowing the process of the next layer to start before finishing the previous layer. This results in reduced latency. Our benchmarking results demonstrate two orders of magnitude reduction in memory consumption using the event-driven depth-first convolution.

\subsection{Hard attention}

Processing high-resolution event-based cameras in neuromorphic platforms is challenging due to the size of neural networks and the excessive amount of operations which is required to finish a task.

One method for processing high-resolution images efficiently is called hard attention \cite{papadopoulos2021hard}. In this approach, a smaller version of the input is processed in a neural network to identify the Regions of Interest (RoI). Then, in parallel neural network(s), these RoIs are zoomed in (using the original high-resolution image) and analyzed to determine the content of the RoI. Implementing such a neural network requires significant pre- and post-processing mechanisms, which can be computationally heavy. To address this issue, we utilized the flexibility of SENECA and implemented a full hard attention solution in the platform. Our experiments with a neuromorphic benchmark (IBM Hand Gesture Recognition) showed that hard attention can improve the energy consumption of the application by 2x and the latency by 3x without sacrificing accuracy. The detailed results of this work are currently under review and will be published soon. 

\section{And more}

The SENECA platform was created as a test bed for exploring and innovating in the field of neuromorphic (both hardware and software) technologies. In addition to what has been mentioned in this paper, many other explorations have been undertaken or are ongoing. For example, using new memory technologies, mapping neural networks with shared memory, exploring 3D architecture, and in-memory and in-material processing.

We have identified certain challenges with the architecture that need to be addressed. One significant challenge arises when using small-scale neural networks, where the incoming event only updates a few neurons. In this case, the number of operations in NPEs is not significant enough to compensate for the overheads. In this case, the per-event preprocessing in RISC-V takes more time than the task's execution. This pushes us back to the domain where RISC-V was the main bottleneck of energy and latency. 

This situation becomes more severe when using depth-first convolution, as it requires more complex preprocessing. Therefore, we have observed that for CNN layers with small channel sizes, a considerable amount of time/energy goes to the RISC-V during inference. This issue can be resolved by introducing specialized depth-first accelerators. 

Additionally, new accelerators need to be designed to handle operations for on-device learning and new types of neural networks, such as transformers. Exploiting weight sparsity is also an ongoing work that needs to be added to SENECA. Weight sparsity exploitation is not trivial when using a simple synchronous vector of NPEs.

\section{Lessons learned and design guidelines}

The SENECA project has progressed from a simple array of RISC-V cores with a NoC to a richer architecture that includes Neural Processing Elements (NPEs), a loop controller, and several software optimizations. In this section, we distill the main lessons learned into a set of design guidelines for future digital neuromorphic processors.

\subsection{Core and NoC design}

\begin{itemize}
  \item \textbf{Start from a flexible core.} A tiny general-purpose core (e.g., RISC-V) provides a fast path to a working neuromorphic platform and makes it easy to explore new neuron models, network topologies, and learning rules. Even if such a core is not optimal for a specific benchmark, it is an excellent vehicle for architectural exploration.
  \item \textbf{Keep the NoC simple.} For the workloads we considered, the NoC was never a performance or energy bottleneck. Event sparsity and high operation density per spike imply that the cost of moving a packet is much smaller than the cost of processing it. A simple mesh with source-based routing and modest routing tables was sufficient in our experiments.
  \item \textbf{Exploit time-multiplexing judiciously.} Time-multiplexing many logical neurons onto one physical core improves area efficiency, but reduces performance and energy efficiency. Designers should tune the time-multiplexing ratio to balance silicon area, throughput, and latency for their target application.
\end{itemize}

\subsection{Hardware acceleration}

\begin{itemize}
  \item \textbf{Add accelerators incrementally.} Comparing SENECA(V1) with Loihi and SpiNNaker2 showed that dedicated neural accelerators can dramatically improve performance and energy efficiency, especially for dense workloads such as fully connected and convolutional layers. Adding NPEs in SENECA(V2) closed most of the gap while preserving programmability.
  \item \textbf{Separate control and computation.} Introducing a loop controller to orchestrate the NPEs reduced instruction-fetch and control overhead on the RISC-V core, improving system efficiency. A hierarchical control structure (RISC-V $\rightarrow$ loop controller $\rightarrow$ NPEs) is a practical pattern for future designs.
  \item \textbf{Target accelerators to dominant kernels.} In our benchmarks, vector-matrix and convolution operations dominated the computation. Focusing accelerators on these kernels provided the largest benefits, while leaving pre- and post-processing on the general-purpose core.
\end{itemize}

\subsection{Software and mapping}

\begin{itemize}
  \item \textbf{Exploit event-driven sparsity in software.} Techniques such as spike grouping can significantly reduce control overhead for fully connected layers by processing multiple events together on the same data. These optimizations are often available ``for free'' once the architecture is flexible enough.
  \item \textbf{Use depth-first convolution to reduce memory footprint.} Event-driven depth-first convolution allows processing CNNs with substantially lower memory requirements than conventional layer-by-layer schemes. In our experiments, this led to orders-of-magnitude reductions in activation memory, which is critical for on-chip deployment.
  \item \textbf{Leverage architectural flexibility for complex models.} Implementing hard-attention style networks for high-resolution event-based vision required substantial pre- and post-processing around the core CNN. The flexibility of the SENECA architecture made it possible to integrate the entire pipeline, leading to both energy and latency improvements without sacrificing accuracy.
\end{itemize}

\subsection{Open challenges}

\begin{itemize}
  \item \textbf{Efficiency for small workloads.} For small neural networks or CNN layers with few channels, the per-event preprocessing on the RISC-V core can dominate the total cost, even in the presence of NPEs. Dedicated accelerators for depth-first convolution and other control-heavy patterns are a promising direction.
  \item \textbf{Support for new model families.} Extending digital neuromorphic processors to on-device learning and transformer-style architectures will likely require new accelerator blocks and execution models that go beyond the current focus on spiking and convolutional networks.
  \item \textbf{Weight sparsity and advanced memories.} Exploiting weight sparsity efficiently with synchronous vectors of NPEs remains non-trivial and is an active topic in SENECA. Integrating new memory technologies and 3D architectures offers additional opportunities to reduce data movement and further improve energy efficiency.
\end{itemize}

These guidelines are not exhaustive, but they summarize the main architectural insights gained from building and iterating on the SENECA platform. We hope they can serve as a practical starting point for students and designers who wish to develop their own digital neuromorphic processors.

\section*{Acknowledgments}
The results used in this tutorial were extracted from several published papers during the SENECA project. They are the result of the hard work of many colleagues, students and co-authors. 

\bibliographystyle{unsrt}  
\bibliography{references}

@inproceedings{sun2023analog,
  title={Analog or Digital In-Memory Computing? Benchmarking Through Quantitative Modeling},
  author={Sun, Jiacong and Houshmand, Pouya and Verhelst, Marian},
  booktitle={2023 IEEE/ACM International Conference on Computer Aided Design (ICCAD)},
  pages={1--9},
  year={2023},
  organization={IEEE}
}

@article{furber2014spinnaker,
  title={The spinnaker project},
  author={Furber, Steve B and Galluppi, Francesco and Temple, Steve and Plana, Luis A},
  journal={Proceedings of the IEEE},
  volume={102},
  number={5},
  pages={652--665},
  year={2014},
  publisher={IEEE}
}

@article{mayr2019spinnaker,
  title={Spinnaker 2: A 10 million core processor system for brain simulation and machine learning},
  author={Mayr, Christian and Hoeppner, Sebastian and Furber, Steve},
  journal={arXiv preprint arXiv:1911.02385},
  year={2019}
}

@article{davies2018loihi,
  title={Loihi: A neuromorphic manycore processor with on-chip learning},
  author={Davies, Mike and Srinivasa, Narayan and Lin, Tsung-Han and Chinya, Gautham and Cao, Yongqiang and Choday, Sri Harsha and Dimou, Georgios and Joshi, Prasad and Imam, Nabil and Jain, Shweta and others},
  journal={Ieee Micro},
  volume={38},
  number={1},
  pages={82--99},
  year={2018},
  publisher={IEEE}
}

@inproceedings{orchard2021efficient,
  title={Efficient neuromorphic signal processing with loihi 2},
  author={Orchard, Garrick and Frady, E Paxon and Rubin, Daniel Ben Dayan and Sanborn, Sophia and Shrestha, Sumit Bam and Sommer, Friedrich T and Davies, Mike},
  booktitle={2021 IEEE Workshop on Signal Processing Systems (SiPS)},
  pages={254--259},
  year={2021},
  organization={IEEE}
}

@article{akopyan2015truenorth,
  title={Truenorth: Design and tool flow of a 65 mw 1 million neuron programmable neurosynaptic chip},
  author={Akopyan, Filipp and Sawada, Jun and Cassidy, Andrew and Alvarez-Icaza, Rodrigo and Arthur, John and Merolla, Paul and Imam, Nabil and Nakamura, Yutaka and Datta, Pallab and Nam, Gi-Joon and others},
  journal={IEEE transactions on computer-aided design of integrated circuits and systems},
  volume={34},
  number={10},
  pages={1537--1557},
  year={2015},
  publisher={IEEE}
}

@article{quian2010measuring,
  title={Measuring sparseness in the brain: comment on Bowers (2009).},
  author={Quian Quiroga, Rodrigo and Kreiman, Gabriel},
  year={2010},
  publisher={American Psychological Association}
}

@article{indiveri2015memory,
  title={Memory and information processing in neuromorphic systems},
  author={Indiveri, Giacomo and Liu, Shih-Chii},
  journal={Proceedings of the IEEE},
  volume={103},
  number={8},
  pages={1379--1397},
  year={2015},
  publisher={IEEE}
}

@article{christensen20222022,
  title={2022 roadmap on neuromorphic computing and engineering},
  author={Christensen, Dennis V and Dittmann, Regina and Linares-Barranco, Bernabe and Sebastian, Abu and Le Gallo, Manuel and Redaelli, Andrea and Slesazeck, Stefan and Mikolajick, Thomas and Spiga, Sabina and Menzel, Stephan and others},
  journal={Neuromorphic Computing and Engineering},
  volume={2},
  number={2},
  pages={022501},
  year={2022},
  publisher={IOP Publishing}
}

@article{schuman2022opportunities,
  title={Opportunities for neuromorphic computing algorithms and applications},
  author={Schuman, Catherine D and Kulkarni, Shruti R and Parsa, Maryam and Mitchell, J Parker and Kay, Bill and others},
  journal={Nature Computational Science},
  volume={2},
  number={1},
  pages={10--19},
  year={2022},
  publisher={Nature Publishing Group}
}

@article{furber2023digital,
  title={Digital neuromorphic technology: current and future prospects},
  author={Furber, Steve},
  journal={National Science Review},
  pages={nwad283},
  year={2023},
  publisher={Oxford University Press}
}

@article{le202364,
  title={A 64-core mixed-signal in-memory compute chip based on phase-change memory for deep neural network inference},
  author={Le Gallo, Manuel and Khaddam-Aljameh, Riduan and Stanisavljevic, Milos and Vasilopoulos, Athanasios and Kersting, Benedikt and Dazzi, Martino and Karunaratne, Geethan and Br{\"a}ndli, Matthias and Singh, Abhairaj and Mueller, Silvia M and others},
  journal={Nature Electronics},
  volume={6},
  number={9},
  pages={680--693},
  year={2023},
  publisher={Nature Publishing Group UK London}
}

@article{modha2023neural,
  title={Neural inference at the frontier of energy, space, and time},
  author={Modha, Dharmendra S and Akopyan, Filipp and Andreopoulos, Alexander and Appuswamy, Rathinakumar and Arthur, John V and Cassidy, Andrew S and Datta, Pallab and DeBole, Michael V and Esser, Steven K and Otero, Carlos Ortega and others},
  journal={Science},
  volume={382},
  number={6668},
  pages={329--335},
  year={2023},
  publisher={American Association for the Advancement of Science}
}

@inproceedings{moreira2020neuronflow,
  title={NeuronFlow: a neuromorphic processor architecture for live AI applications},
  author={Moreira, Orlando and Yousefzadeh, Amirreza and Chersi, Fabian and Cinserin, Gokturk and Zwartenkot, Rik-Jan and Kapoor, Ajay and Qiao, Peng and Kievits, Peter and Khoei, Mina and Rouillard, Louis and others},
  booktitle={2020 Design, Automation \& Test in Europe Conference \& Exhibition (DATE)},
  pages={840--845},
  year={2020},
  organization={IEEE}
}

@article{richter2023speck,
  title={Speck: A Smart event-based Vision Sensor with a low latency 327K Neuron Convolutional Neuronal Network Processing Pipeline},
  author={Richter, Ole and Xing, Yannan and De Marchi, Michele and Nielsen, Carsten and Katsimpris, Merkourios and Cattaneo, Roberto and Ren, Yudi and Liu, Qian and Sheik, Sadique and Demirci, Tugba and others},
  journal={arXiv preprint arXiv:2304.06793},
  year={2023}
}

@article{demler2019brainchip,
  title={Brainchip akida is a fast learner, spiking-neural-network processor identifies patterns in unlabeled data},
  author={Demler, Mike},
  journal={Microprocessor Report},
  year={2019}
}

@article{stuijt2021mubrain,
  title={$\mu$Brain: An event-driven and fully synthesizable architecture for spiking neural networks},
  author={Stuijt, Jan and Sifalakis, Manolis and Yousefzadeh, Amirreza and Corradi, Federico},
  journal={Frontiers in neuroscience},
  volume={15},
  pages={664208},
  year={2021},
  publisher={Frontiers}
}

@inproceedings{yousefzadeh2022seneca,
  title={SENeCA: Scalable energy-efficient neuromorphic computer architecture},
  author={Yousefzadeh, Amirreza and Van Schaik, Gert-Jan and Tahghighi, Mohammad and Detterer, Paul and Traferro, Stefano and Hijdra, Martijn and Stuijt, Jan and Corradi, Federico and Sifalakis, Manolis and Konijnenburg, Mario},
  booktitle={2022 IEEE 4th International Conference on Artificial Intelligence Circuits and Systems (AICAS)},
  pages={371--374},
  year={2022},
  organization={IEEE}
}

@incollection{molendijk2022benchmarking,
  title={Benchmarking the Epiphany processor as a reference neuromorphic architecture},
  author={Molendijk, Maarten and Vadivel, Kanishkan and Corradi, Federico and van Schaik, Gert-Jan and Yousefzadeh, Amirreza and Corporaal, Henk},
  booktitle={Industrial Artificial Intelligence Technologies and Applications},
  pages={21--34},
  year={2022}
}

@article{yousefzadeh2022energy,
  title={Energy-efficient in-memory address calculation},
  author={Yousefzadeh, Amirreza and Stuijt, Jan and Hijdra, Martijn and Liu, Hsiao-Hsuan and Gebregiorgis, Anteneh and Singh, Abhairaj and Hamdioui, Said and Catthoor, Francky},
  journal={ACM Transactions on Architecture and Code Optimization (TACO)},
  volume={19},
  number={4},
  pages={1--16},
  year={2022},
  publisher={ACM New York, NY}
}

@article{tang2023seneca,
  title={SENECA: building a fully digital neuromorphic processor, design trade-offs and challenges},
  author={Tang, Guangzhi and Vadivel, Kanishkan and Xu, Yingfu and Bilgic, Refik and Shidqi, Kevin and Detterer, Paul and Traferro, Stefano and Konijnenburg, Mario and Sifalakis, Manolis and van Schaik, Gert-Jan and others},
  journal={Frontiers in Neuroscience},
  volume={17},
  year={2023},
  publisher={Frontiers Media SA}
}

@article{tang2023open,
  title={Open the box of digital neuromorphic processor: Towards effective algorithm-hardware co-design},
  author={Tang, Guangzhi and Safa, Ali and Shidqi, Kevin and Detterer, Paul and Traferro, Stefano and Konijnenburg, Mario and Sifalakis, Manolis and van Schaik, Gert-Jan and Yousefzadeh, Amirreza},
  journal={arXiv preprint arXiv:2303.15224},
  year={2023}
}

@article{kevin_paper,
  title={Optimizing Event-Based Neural Networks on Digital Neuromorphic Architecture: A Comprehensive Design Space Exploration},
  author={Xu, Yingfu and et al},
  journal={Frontiers in Neuroscience},
  year={2024},
  publisher={Frontiers Media SA}
}

@misc{TEMPO,
  title = {Technology and hardware for neuromorphic computing},
  howpublished = {\url{https://cordis.europa.eu/project/id/826655}}
}

@misc{ANDANTE,
  title = {Ai for New Devices And Technologies at the Edge},
  howpublished = {\url{https://cordis.europa.eu/project/id/876925}}
}

@misc{MNEMOSENE,
  title = {Computation-in-memory architecture based on resistive devices},
  howpublished = {\url{https://cordis.europa.eu/project/id/780215}}
}

@misc{DAIS,
  title = {Distributed Artificial Intelligent Systems},
  howpublished = {\url{https://cordis.europa.eu/project/id/101007273}}
}

@misc{MeM-Scales,
  title = {Memory technologies with multi-scale time constants for neuromorphic architectures},
  howpublished = {\url{https://cordis.europa.eu/project/id/871371}}
}

@misc{REBECCA,
  title = {Reconfigurable Heterogeneous Highly Parallel Processing Platform for safe and secure AI},
  howpublished = {\url{https://cordis.europa.eu/project/id/101097224}}
}

@misc{NEUROKIT2E,
  title = {Open source deep learning platform dedicated to Embedded hardware and Europe},
  howpublished = {\url{https://cordis.europa.eu/project/id/101112268}}
}

@misc{NimbleAI,
  title = {ULTRA-ENERGY EFFICIENT AND SECURE NEUROMORPHIC SENSING AND PROCESSING AT THE ENDPOINT},
  howpublished = {\url{https://cordis.europa.eu/project/id/101070679}}
}

@misc{NeuralMagic,
  title = {Neural Magic Scales up MLPerf™ Inference v3.0 Performance With Demonstrated Power Efficiency; No GPUs Needed},
  howpublished = {\url{https://neuralmagic.com/blog/neural-magic-scales-up-mlperf-inference-performance-with-demonstrated-power-efficiency-no-gpus-needed/}}
}

@inproceedings{yousefzadeh2018performance,
  title={Performance comparison of time-step-driven versus event-driven neural state update approaches in spinnaker},
  author={Yousefzadeh, Amirreza and Soto, Mikel and Serrano-Gotarredona, Teresa and Galluppi, Francesco and Plana, Luis and Furber, Steve and Linares-Barranco, Bernabe},
  booktitle={2018 IEEE International Symposium on Circuits and Systems (ISCAS)},
  pages={1--4},
  year={2018},
  organization={IEEE}
}

@article{rostami2022prop,
  title={E-prop on SpiNNaker 2: Exploring online learning in spiking RNNs on neuromorphic hardware},
  author={Rostami, Amirhossein and Vogginger, Bernhard and Yan, Yexin and Mayr, Christian G},
  journal={Frontiers in Neuroscience},
  volume={16},
  pages={1018006},
  year={2022},
  publisher={Frontiers}
}

@inproceedings{kelber2020mapping,
  title={Mapping deep neural networks on SpiNNaker2},
  author={Kelber, Florian and Wu, Binyi and Vogginger, Bernhard and Partzsch, Johannes and Liu, Chen and Stolba, Marco and Mayr, Christian},
  booktitle={Proceedings of the 2020 Annual Neuro-Inspired Computational Elements Workshop},
  pages={1--3},
  year={2020}
}

@article{knight2018gpus,
  title={GPUs outperform current HPC and neuromorphic solutions in terms of speed and energy when simulating a highly-connected cortical model},
  author={Knight, James C and Nowotny, Thomas},
  journal={Frontiers in neuroscience},
  volume={12},
  pages={941},
  year={2018},
  publisher={Frontiers}
}

@article{chen2020slide,
  title={Slide: In defense of smart algorithms over hardware acceleration for large-scale deep learning systems},
  author={Chen, Beidi and Medini, Tharun and Farwell, James and Tai, Charlie and Shrivastava, Anshumali and others},
  journal={Proceedings of Machine Learning and Systems},
  volume={2},
  pages={291--306},
  year={2020}
}

@article{shukla2019remodel,
  title={REMODEL: Rethinking deep CNN models to detect and count on a NeuroSynaptic system},
  author={Shukla, Rohit and Lipasti, Mikko and Van Essen, Brian and Moody, Adam and Maruyama, Naoya},
  journal={Frontiers in neuroscience},
  volume={13},
  pages={4},
  year={2019},
  publisher={Frontiers Media SA}
}

@misc{RaveNoC,
  title = {RaveNoC - configurable Network-on-Chip},
  howpublished = {\url{https://github.com/aignacio/ravenoc}}
}

@article{papamichael2015connect,
  title={The CONNECT network-on-chip generator},
  author={Papamichael, Michael K and Hoe, James C},
  journal={Computer},
  volume={48},
  number={12},
  pages={72--79},
  year={2015},
  publisher={IEEE}
}

@misc{AwesomeMesh,
  title = {Awesome-Mesh},
  howpublished = {\url{https://github.com/moarpepes/awesome-mesh}}
}

@inproceedings{blouw2019benchmarking,
  title={Benchmarking keyword spotting efficiency on neuromorphic hardware},
  author={Blouw, Peter and Choo, Xuan and Hunsberger, Eric and Eliasmith, Chris},
  booktitle={Proceedings of the 7th annual neuro-inspired computational elements workshop},
  pages={1--8},
  year={2019}
}

@article{yan2021comparing,
  title={Comparing Loihi with a SpiNNaker 2 prototype on low-latency keyword spotting and adaptive robotic control},
  author={Yan, Yexin and Stewart, Terrence C and Choo, Xuan and Vogginger, Bernhard and Partzsch, Johannes and H{\"o}ppner, Sebastian and Kelber, Florian and Eliasmith, Chris and Furber, Steve and Mayr, Christian},
  journal={Neuromorphic Computing and Engineering},
  volume={1},
  number={1},
  pages={014002},
  year={2021},
  publisher={IOP Publishing}
}

@article{papadopoulos2021hard,
  title={Hard-attention for scalable image classification},
  author={Papadopoulos, Athanasios and Korus, Pawel and Memon, Nasir},
  journal={Advances in Neural Information Processing Systems},
  volume={34},
  pages={14694--14707},
  year={2021}
}

@article{yousefzadeh2017multiple,
  title={On multiple AER handshaking channels over high-speed bit-serial bidirectional LVDS links with flow-control and clock-correction on commercial FPGAs for scalable neuromorphic systems},
  author={Yousefzadeh, Amirreza and Jab{\l}o{\'n}ski, Miros{\l}aw and Iakymchuk, Taras and Linares-Barranco, Alejandro and Rosado, Alfredo and Plana, Luis A and Temple, Steve and Serrano-Gotarredona, Teresa and Furber, Steve B and Linares-Barranco, Bernab{\'e}},
  journal={IEEE transactions on biomedical circuits and systems},
  volume={11},
  number={5},
  pages={1133--1147},
  year={2017},
  publisher={IEEE}
}

\end{document}